# A Note on the Origin of Inertia

A. Schlatter; R.E. Kastner

The Quantum Institute, Gloversville, NY 12078, USA; schlatter.a@bluewin.ch
Department of Philosophy, University of Maryland, College Park, MD 20740, USA; rkastner@umd.edu

**Abstract:** The question of where the inertial properties of matter come from has been open for a long time. Isaac Newton considered inertia an intrinsic property of matter. Ernst Mach held a different view whereby the inertia of a body comes from its interaction with the rest of the universe. This idea is known today as Mach's principle. We discuss Mach's principle based on transactional gravity, the recently developed completion of the entropic gravity program by the physics of quantum-events, induced by transactions [1,2]. A consequence of the analysis is a fundamental relation between the gravitational constant $G$ and the total mass in the causal universe, derived by means of entropic principles.

**Keywords:** transaction; space-time; entropic gravity; inertial frame; Mach's principle.

## 1. Introduction and background

In context of the fameous bucket experiment, Newton concluded that acceleration is only definable relative to absolute space and that inertial reference frames, in which his laws are valid, are consequently also determined by this abstract notion. In this view inertia is an intrinsic property of matter. Absolute space, however, is not observable and therefore there arose opposition to Newton's view early on. For Leibniz, who thought of space (and time) as a relational property of matter, acceleration of a material body could only be defined relative to other material systems. Mach [3] was more concrete and thought that acceleration was in principle defined relative to all the matter in the universe and that inertial frames are those unaccelerated relative to "the fixed stars", a view which is today called Mach's principle.[1] If the rest of the universe determines the inertial frames, then inertia is not an intrinsic property of matter, but is a result of the interaction of matter with the rest of the universe. The observational fact that gravitation is locally indistinguishable from an inertial force, because each induces the same acceleration in all bodies, leads to the insight that inertial and gravitational mass, i.e. the "charge" of gravity, are equal. Consequently, Einstein thought that it is gravitational interaction with the whole universe which gives rise to a body's inertia and he intended to build Mach's principle into his theory of gravity.

The fact that Newton's laws locally hold so accurately without the slightest reference to the rest of the universe made it hard to accept Mach's ideas in the first place and the fact that Einstein himself [5] demonstrated that in his theory a body moving in otherwise empty space[2] has inertial properties, made Einstein's attempt to implement Mach's principle look a partial success at most. Sciama [6] developed a theory of inertia based on a gravitational potential of the universe and a corresponding vector potential satisfying Maxwell-type equations. In case of rectilinear and rotational motion he obtained a combination of Newton's laws of motion and of gravitation, with the inertial frame and fictious force determined by Mach's principle. While the model does not explain all motion, it particularly makes some assumptions regarding the form of the potential, which need a firmer basis. Other theories of inertia have been developed over time, see e.g. [7,8,9].

Recently, it was shown that the physics of quantum-events, induced by transactions, does complete the entropic gravity program [1,2,10]. We will call this theory "transactional gravity". In this paper we discuss Mach's principle in light of the insights of transactional gravity. We will find deeper reasons for some just empirically hardened properties of gravity and give support for the validity of Mach's principle in general relativity by deriving the form of the potential of the universe in Sciama's model and excluding certain spacetime solutions on physical ground. We also derive from entropic principles, independently of any model, a definition of the gravitational constant by the total mass in the causal universe.

## 2. Gravitational force from entropic considerations

In this section we give a concise introduction to transactional gravity in three steps. First, we introduce the notion of transactions and second, we will apply their entropic consequences in order to derive the gravitational force. We will then sketch how the field theory of gravity and Einstein's equations are derived. For a detailed presentation, especially of the last part, we refer to [1,2].

---

[1] There are number of techically distinct formulations of Mach's principle [4].
[2] Minkowski space is a solution of Einstein's vacuum equations.

### 2.1 Spacetime and transactions

Quantum amplitudes of closed, isolated systems are represented as unit vectors in a Hilbert space, $\psi_x \epsilon H$, also called quantum states. In the transactional interpretation [11] a quantum state $\psi_x$ is launched as an "offer-wave" by an emitter and gets possible responses by "confirmation-waves", represented by dual vectors $\psi_y^*$ launched by possible absorbers. The selection of a specific "response" $\psi_x^*$ is fundamentally indeterministic and leads to a "transaction", which is the actualization of absorption and emission as real events in spacetime, and whose probability (density) is $\left| \delta_{(y-x)} * \psi_x \right|^2 = \left| \psi_y \right|^2$. The relativistic transactional interpretation [11] offers additionally the reason why offer-waves (and confirmation-waves) are actually being created, by focusing on the electromagnetic interaction. Relativistic electromagnetic interactions can be thought of as the mutual exchange of virtual photons by quantum fields, creating possibilities in a pre-spacetime process. Transactions, in turn, are characterized by the exchange of real photons and their four-momenta between emitter and absorber. While virtual photons correspond to the Coulomb force and such interactions are unitary, real photons correspond to radiative processes, which are non-unitary interactions ([11], Chapter 5). The general amplitude for emission and absorption of real photons is the coupling amplitude between matter-and gauge fields, and the non-unitary transactional process can arise if the conservation laws are satisfied. By this non-unitary exchange of four-momentum the quantum states of emitter and absorber collapse, and the physical systems are localized at the corresponding spacetime points (regions).

Empirical spacetime thus becomes the connected set of emission-and absorption points, between which space-time (null)intervals are being created through the four-momentum of the exchanged photons. It is here, where the transactional view touches causal-set theory [12], in which events spread in spacetime by a stochastic Poisson-process. Boson exchange, understood as a decay-process in quantum field theory, is then a special case in this model.[3] Note, that the actualization of a spacetime interval amounts to spontaneously breaking the unitary evolution of the quantum states. At the same time, the exchanged four-momentum selects a space-direction, whereas a time-direction is a priori determined, since only positive energy is being transferred.[4] Because there is no preferred emission-direction, the process is spatially isotropic, and because the whole mechanism is indifferent to specific locations, it is also homogeneous. Observed inhomogeneities of the universe are then the consequence of a possible anisotropic distribution of initial transactions.

So far, we have motivated the idea that the formalism of quantum physics is suited to explain the emergence of empirical space and time as unified, yet distinct, dimensions by the mechanism of transactions. Real photon exchange creates metric relations between emitters and absorbers, as described in [2], and the mechanism contains therefore an intrinsic way to measure time-intervals by means of the exchanged photons. This approach clearly lends itself to the relational view of spacetime as it emerges from pairs of emitters and absorbers, and there is no spacetime without matter (although the matter itself is not a component of metrical spacetime; [11], Chapter 8). Mathematically, quantum states can be described as fields parametrized by spacetime coordinates. Note that there is no circularity here and this description does not suggest that spacetime has a real a priori existence of its own. It is only a model representation of our observations, where we never measure standalone spacetime points. Hence the term "empirical spacetime" in order to distinguish observed reality from mathematical models. We also note that the conception of quantum states as fields over spacetime uses the spacetime parameters as possibilities for localization relative to a particular inertial frame, and that such quantum states are not physically 'in spacetime' where the latter is understood as the emergent manifold of connected events. In this regard it is worth quoting our remarks from [2] concerning this point:

"It is also important to note that the events are not to be identified with the material systems (e.g., atoms) giving rise to them. That is, the systems themselves are never part of the spacetime manifold nor are they within it. It is only their *activities*--the events--that are elements of the spacetime manifold, as are the photonic connections between them, which establish the structure of the manifold. In common parlance one refers to a spacetime point being "occupied" by matter, but in our picture that is really a shorthand for the idea that a material system has become associated with that point (region) (understood as simply an element of a particular reference frame) via an actualized event. (Of course, given the spreading of wave packets, such a system will not remain indefinitely localized.) "

---

[3] The transactional interpretation thinks differently of spacetime than the causal-set approach does, which is unimportant for our purposes.
[4] This amounts to the choice of the Feynmann-propagator as opposed to the Dyson-propagator.

It is now important to note that absorption of a photon and the corresponding localization of a physical system has implications for the entropy balance. The calculation of this entropy is key to derive gravitational acceleration and hence to complete an idea of Verlinde [10]. Note also that, because we work with the electromagnetic interaction, the mass of material systems is a priori the inertial (rest)mass, which we measure in the laboratory and the existence of inertia is an empirical fact. The remainder of this section is a recap of material in [2], provided for background purposes.

### 2.2 Spatial information

We assume to work in a local inertial frame throughout the following exposition. Let there be a bounded region $\Omega \subset \mathbb{R}^{3N}$[5], $N \epsilon \mathbb{N}$, on a spatial hyperplane and a partition by balls

$$\mathcal{B} = \left\{ B_{\varepsilon_n}(x_n) \right\}_{x_n \epsilon \Omega, \varepsilon_n > 0}, \bigcup_{x_n} B_{\varepsilon_n}(x_n) = \Omega. \tag{1}$$

Relative to the partition $\mathcal{B}$, a position-information can be attributed to a quantum system in terms of its quantum state over $\Omega$, $\psi(x) \epsilon L^2(\Omega)$, by

$$I^{\mathcal{B}}(\psi) = -\sum_{x_n \epsilon \Omega} p_{x_n} ln(p_{x_n}), \qquad p_{x_n} = \int_{B_{\varepsilon_n}(x_n)} |\psi(x)|^2 dx. \tag{2}$$

By multiplication with the Boltzmann constant, $k_B$, we get

$$S^{\mathcal{B}}(\psi) = I^{\mathcal{B}}(\psi) k_B.[6] \tag{3}$$

We can ask, whether it is possible to take a different perspective and attribute information not to material systems, but to regions or, idealized, single points $x_0 \epsilon \mathbb{R}^{3N}$. A point, $x_0 \epsilon \Omega$, can empirically be associated with matter or not and hence represents in this sense one bit of information. Given a physical system, $\psi(x) \epsilon L^2(\Omega)$, we can therefore state that the information of the one bit, $x_0 \epsilon \Omega$, with respect to $\psi(x)$ and the partition $\mathcal{B}$ (1)[7] is

$$I_{\psi}^{\mathcal{B}}(x_0) = -\left[ p_{x_0} ln(p_{x_0}) + (1 - p_{x_0}) ln(1 - p_{x_0}) \right]. \tag{4}$$

To find a generic definition, we have to account for all possible partitions $\mathcal{B}$, which amounts to taking into account all probabilities, $0 \leq p_{x_0} \leq 1$. Since we always find some bounded $\Omega \subset \mathbb{R}^{3N}$ with $x_0 \epsilon \Omega$, we can define the information $I(x_0)$, $x_0 \epsilon \mathbb{R}^{3N}$, by

$$I(x_0) = -2 \int_0^1 p_{x_0} ln(p_{x_0}) dp_{x_0} = \frac{1}{2}. \tag{5}$$

Evidently, (5) is not only independent of a chosen partition $\mathcal{B}$, but also of the particular material system $\psi(x)$. While the choice of a particular $\mathcal{B}$ is, of course, frame-dependent, the described process will lead to the definition of $I(x_0)$ by equation (10) in every local inertial frame. By (3) we arrive at the corresponding one-bit entropy

---

[5] The dimension allows for multi-particle systems.
[6] The identification of information entropy and thermodynamic entropy in case of transactions is justified in [13].
[7] We pick the ball $B_{\varepsilon_n}(x_n)$ with $x_0 \epsilon B_{\varepsilon_n}(x_n)$ and minimal $|x_0 - x_n|$.

$$S(x_0) = I(x_0)k_B = \frac{1}{2}k_B. \tag{6}$$

For reasons, to be justified later, (6) will in 2.4 be rescaled by a factor of $4\pi$ to

$$\tilde{S}(x_0) = 4\pi k_B I(x_0) = 2\pi k_B. \tag{7}$$

### 2.3 Transactions and a holographic principle

Assume there is a transaction between two material systems in physical space $\mathbb{R}^3$[8] through the exchange of a real (on-shell) photon. A transaction breaks the unitary symmetry of the interaction and localizes absorber and emitter. If it takes the photon a time interval $\Delta t_R = \frac{R}{c}$ from the emitter to the absorber, then, next to the time interval $\Delta t_R$, there is the spatial distance $R > 0$ being generated by the process. For symmetry reasons[9], the possible locations of the absorber lie equiprobably on the sphere $S_R$. We can define in analogy to (5) and (6) the total entropy of the sphere $S(S_R)$ around the emitter by

$$S(S_R) = k_B I(S_R) = \frac{k_B}{2}N_R, \tag{8}$$

where $N_R$ is just the number of bits on the sphere. For this number we set, with $l_P = \sqrt{\frac{G\hbar}{c^3}}$ and $A_R = 4\pi R^2$,

$$N_R = \frac{A_R}{l_P^2} = 4\pi\frac{R^2 c^3}{G\hbar}. \tag{9}$$

So we get

$$S(S_R) = k_B\frac{A_R}{2l_P^2}. \tag{10}$$

Note that by (8) equation (9) can be written in a more suggestive form to which we will return later in paragraph 3.3

$$G \cdot I(S_R) = \frac{2\pi R^2 c^3}{\hbar}. \tag{11}$$

Remember that expression (8) is the sum of all entropies $S(x)$, $x \epsilon S_R$, which in turn are calculated over all possible local probabilities, $0 \le p_x \le 1$, $x \epsilon S_R$. So $S(S_R)$ can be considered to be the entropy contained in all possible absorptions on the sphere $S_R$.

Let us assume that in a concrete physical situation the fields defined over the ball $B_R$, $\psi(x)\epsilon L^2(B_R)$, have a total mass $M$. We then know the total energy to be $E_{tot} = Mc^2$ and derive from (10) and the thermodynamic relation $E = S \cdot T$ by the thermodynamic equivalence principle the existence of a formal temperature $T = T(M, R)$ satisfying

$$Mc^2 = S(S_R) \cdot T(M, R) = k_B\frac{A_R}{2l_P^2}T(M, R). \tag{12}$$

By (12) we get for $T(M, R)$, with $g_R = \frac{GM}{R^2}$, the expression

---

[8] We idealize and restrict the argument space of the quantum state $\psi$ by projecting
onto the diagonal $\mathbb{R}^{3N} \to \{\underline{x}\epsilon\mathbb{R}^{3N}|x_1 = x_2 = \cdots x_{3N}\} \approx \mathbb{R}^3$.
[9] Photons are emitted in all spatial directions with equal probability.

$$T(M,R) = \frac{\hbar M G}{2\pi c k_B R^2} = \frac{\hbar g_R}{2\pi c k_B}. \tag{13}$$

We are now ready for the final step.

### 2.4 Entropic force

We are now in a position to reconstruct the argument in [10]. Assume that a particle of mass $m$ enters into a transaction with a system of total mass $M$, $m \ll M$, a process which localizes the particle at a point $x_0 \epsilon S_R$ for some $R > 0$. This process makes position-information (6) available and consequently by the 2nd law, there is a (minimal) amount of entropy $\Delta \tilde{S}(x_0)$ added to the thermal environment on the surface. Since the Planck length is the smallest length possible, we multiply (6) by $4\pi$ and hence set the entropy of the point-particle $\tilde{S}(x_0) = 4\pi k_B I(x_0)$ (7). Due to (10) we therefore have $\tilde{S}(x_0) = S(S_{l_P})$[10]. We further account for the fact that a particle is not really point-like, but still structureless, by using its (reduced) Compton radius $\lambda_C = \frac{\hbar}{mc}$ and postulate that the full information is part of the space-like surface $S_R$ only, if the particle is at a Compton-distance from it and that the information decreases linearly, if it is getting closer to the surface, (i.e. [10], [14])

$$I_{\Delta x}(x_0) = \frac{1}{2}\frac{mc}{\hbar} \cdot \Delta x, 0 \leq \Delta x \leq \lambda_C. \tag{14}$$

Associated with the entropy difference there is by (12) a (minimal) amount of energy $E_{\Delta \tilde{S}(x_0)}$ given by

$$E_{\Delta \tilde{S}(x_0)} = 4\pi k_B I_{\Delta x}(x_0) T(M,R). \tag{15}$$

The definition of energy as work, i.e. force along a path, leads to a corresponding force $F_G$ by

$$E_{\Delta \tilde{S}(x_0)} = F_G \cdot \Delta x. \tag{16}$$

By (13) and (14) we get from (16)

$$F_G \cdot \Delta x = 2\pi k_B \frac{mc}{\hbar} \frac{\hbar M G}{2\pi c k_B R^2} \cdot \Delta x, \tag{17}$$

and finally the expression for $F_G$

$$F_G = m \cdot g(M,R) = m \frac{GM}{R^2}. \tag{18}$$

The force $F_G$ is attractive, since the entropy-gradient $\Delta \tilde{S}(x_0)$ points to the surface $S_R$. The above derivation of $F_G$ has shown that the concept of spatial information and transactions lead, together with the thermodynamic equivalence principle (12), to the existence of gravity as an entropic force, emerging from the coming-into-being of empirical reality (18).[11]

Once we consider that transactions possess by the exchanged photons naturally inbuilt clocks which measure the rhythm of becoming by an invariant gauge $c$, the speed of light, the way is open to derive a metric structure of spacetime, locally governed by Einstein's equations. This reflects the fact that the light-cone structure together with a length-gauge in time-like direction locally determine the four metric [15]. We are not giving the details here, since they can be found in [1,2]. In the process we also find the reason for the general validity of the clock-hypothesis.[12] There is, however, one other important fact which we need to mention. While the energy of the transferred photons implies a natural rhythm of becoming, the transferred three-momentum has a repulsive effect which enters Einstein's equations in form of a scalar Λ which can be interpreted as a cosmological constant. If the number $\varrho_\gamma$ denotes the average number of transactions per spacetime volume, then Λ turns out to be [1,2]

$$\Lambda = 4\pi^2 l_P^2 \varrho_\gamma. \tag{19}$$

---

[10] As explained in 2.1, entropy is understood as unavailable position-information, which is calculated as an average (4,5) over all possible systems. Transactions make this information available, a fact which must be compensated by the second law. In this sense "a particle adds one bit of information to the surface" [10].

[11] We have worked out the minimal acceleration, based on the minimal dissipation in (19). Generally, we can expect $a(M,R) \geq g(M,R)$. This possibility is implicitly responsible for the validity of the more general Einstein's equations.

[12] Timelike curves on a Lorentz-manifold can be approximated arbitrarily closely by a zig-zag of null curves (i.e. light clocks) [16].

### 3. Mach's principle

So far, we have derived gravity from the theory of transactions and the corresponding processing of information. As a consequence, properties of gravity, so far only of empirical nature, gain a deeper explanation.

#### 3.1 The nature of gravity

Based on transactions, the elements of empirical spacetime come into being in a homogeneous and isotropic fashion as a causal set of emitters and absorbers.[13] A natural direction of time is defined by the transport of positive energy from emitter to absorber. This directly leads, as shown above, to an attractive entropic force equivalent, as it turns out, to gravity whose "charge" is inertial mass. Gravity is therefore universal and connot be screened off. This attractive force is exercised on a massive system at a point in spacetime by all the masses from which it absorbs a real photon at that point. Hence the force propagates at the speed of light. The fact that four-momenta are exchanged and that the "charge" is inertial mass, allows the geometrization of gravity, whereby the light cone structure and the energy-component lead to the field equations and the three-component induces a cosmological constant Λ. The fact that Bergson found so hard to believe [17], namely that light clocks should universally govern the duration of every process in the empirical world, gets thus a firm foundation. We also realize that gravity is not a separate quantum field to be unified with electrodynamics in the conventional manner, but is instead an immediate consequence of radiation and the corresponding information processing.

We now know what the nature of gravity is and that gravitational mass must be equal to inertial mass. But is inertial mass an inherent property of matter or is it in turn the result of gravity? We are going to dicuss this question in light of the findings of transactional gravity and Sciama's model.

#### 3.2 Mach's principle

In the theory of transactional gravity there is no absolute space and time. In fact, there is no spacetime without matter. Therefore, acceleration cannot be explained relative to these abstract notions. By the same token, Einstein's doubt whether general relativity implements Mach's principle, because of the existence and properties of Minkowski spacetime, seems unfounded. Neither empty space nor a single physical system in otherwise empty space can exist, because transactions come in pairs of emitter and absorber. Thus, Minkowski spacetime is seen as an unphysical idealization. Empirical spacetime and gravity are inextricably intertwined, since both are consequences of transactions. While quantum fields are not elements of the empirical realm, [14] this realm is the only place where we can measure inertial mass. Sciama realized a model [6] to implement Mach's principle considering the behaviour of a test-particle[15] in presence of a single body superposed on the smoothed out universe. He explicitely modeled the impact of the universe on the test-particle by introducing a potential $\Phi_U$ and a corresponding vector potential $\underline{A}$. As it turns out, the properties of transactional gravity determine the form of this potential.

Since transactions are equiprobable in all directions and do not depend upon the point of emission, the average mass-density $\rho_U$ of the empirical universe can be assumed homogeneous and isotropic at large scales. By the same reason, the total mass at a given distance $R$ from the test-particle increases roughly with the square of that distance $R^2$. The presence of $\Lambda$ (19) implies that the universe expands and empirically there holds Hubble's relation between the expansion velocity $v$ and the distance from the test particle $R$

$$v = H_0 \cdot R, \tag{20}$$

where $H_0$ is Hubble's parameter. We can further assume that the universe is approximately flat, fitting the observed data. At each point there is then a natural state of rest relative to the universe, namely, that in which the observed red-shifts in all directions are equal. In order to derive inertial movement, Sciama postulates that in the rest-frame of any test-particle the total gravitational field at the particle arising from all the other matter in the universe is zero. Since, as we have seen in paragraph 3.1, gravity propagates at the speed of light, only the region within the Hubble-radius $R_U = c/H_0$, i.e. the causal universe, has an impact.[16] Combining all this with (18), there must hold

$$\Phi_U = -\int_{B_{R_U}} \frac{\varrho_U}{R} dV = -2\pi\varrho_U \left(\frac{c}{H_0}\right)^2. \tag{21}$$

---

[13] Note that this fromulation does not assume the pre-existence of space. Space points in three dimensions are generated by transactions which do not distinguish direction or location. Hence the symmetry of space.

[14] Maybe this is the deeper reason why in quantum field theories the values of the particle-masses have to be plugged in "by hand". QFT does not include gravity.

[15] With unit mass.

The same result is true, if the test-particle moves relative to the universe with a small velocity $-\underline{v}(t)$. By defining a vector potential

$$\underline{A} = -\int_{B_{R_U}} \frac{\varrho_U \underline{v}}{cR} dV = \frac{\Phi_U}{c} \underline{v}(t), \tag{22}$$

Sciama gets for the gravito-electric field

$$\underline{E}_g = -\nabla\Phi - \frac{1}{c}\frac{\partial \underline{A}}{\partial t} = -\frac{\Phi_U}{c^2}\frac{\partial \underline{v}}{\partial t}. \tag{23}$$

He then superposes a body of mass $M$ to this universe plus test-particle at distance $R$ in unit-direction $\hat{\underline{R}}$ relative to the test-particle with potential $\phi = -M/R$. In analogy to equation (23) its respective gravito-electric field is

$$\underline{e}_g = -\frac{M}{R^2}\hat{\underline{R}} - \frac{\phi}{c^2}\frac{\partial \underline{v}}{\partial t}. \tag{24}$$

With $\hat{\underline{R}} \cdot (d\underline{v}/dt) = dv/dt$, the total field at the test particle is zero, $(\underline{E}_g + \underline{e}_g) \cdot \hat{\underline{R}} = 0$, if

$$\frac{M}{R^2} = -\left(\frac{\Phi_U + \phi}{c^2}\right)\frac{dv}{dt}. \tag{24}$$

Finally, Sciama sets $G = -\left(\frac{c^2}{\Phi_U + \phi}\right)$ and gets Newton's inertial form of gravity. With $\phi \ll \Phi_U$ there holds $G\Phi_U \approx -c^2$ and by (21) with $M_U$ denoting the mass of the universe

$$G \sim c^2 \frac{R_U}{M_U}. \tag{25}$$

The fact that nearby matter of the smoothed universe does not noticebly disturb inertial frames is due to the fact that with increasing distance the influence of matter (21) drops more slowly than the inverse of its mass increase. In case of uniform rotation of body and universe, Sciama gets by the same reasoning from the gravito-electric field the equation

---

[16] The Hubble sphere with Hubble radius $R_H = c \cdot H_0$ is a good approximation of the causal universe in case of a constant Hubble parameter.
In an accelerated expansion one has to take the particle horizon $R_P = \int_0^T \frac{c}{a(t)} dt$, where $a(t)$ is a time variable scale factor, with $a(T) = 1$, and $T$ is the age of the universe. In the FLRW-model there holds $\dot{a}/a = H_0$. We will just write $R_U$ for the radius of the causal universe in any case.

$$\frac{M}{R^2} = \omega^2 R. \tag{26}$$

In the rest frame of the universe this is simply Newton's equation for circular motion. In the test-particle's rest-frame, however, it amounts to the introduction of a fictious centrifugal force. By the same token he also obtains the Coriolis force $F_C$ from the gravito-magnetic field $\underline{B}$

$$\underline{B} = curl\underline{A} = 2\underline{\omega}\,,\;\; F_C = \underline{v} \times \underline{B} = 2\underline{v} \times \underline{\omega}. \tag{27}$$

Furthermore, it is important to note that Davidson [18] showed that Sciama's model is fully incorporated in general relativity, where it is recovered as a weak-field limit in quasi-flat spacetime, a fact that by the theory of transactional gravity we would, of course, expect.[17]

    Hence, we see that the Sciama's model of a Machian mechanics can be considered as a natural consequence of transactional gravity. Of course, by following rom Sciama's model, the key relation (25) is based on a special case. But indeed, transactional gravity provides more.

### 3.3 The gravitational constant

    Relation (25) is a necessary condition to explain for instance Galilei's free fall experiment in a Machian way. A relation like (25) was already in the minds of Schrödinger [19] and Dirac [20] without proper deduction from a theory, though. It allows to deduce properties of the whole universe by local observation, right in the Machian spirit. Transactional gravity establishes independently of any spacetime model a relation between the constant $G$ and parameters of the causal universe. By equation (11) and the holographic principle [21] we have a definition of $G$ involving the total information $I_U$ and the radius $R_U$

$$G = \frac{2\pi c^3}{\hbar} \frac{R_U^2}{I_U}. \tag{28}$$

In order to express the "mass of information" we use the Bekenstein bound [22], which quantifies the maximal information within a sphere of radius $R$. Within the causal universe it is given by

$$I_U = \frac{S_U}{k_B} = \frac{2\pi R_U E_U}{\hbar c} = \frac{2\pi c R_U M_U}{\hbar}. \tag{29}$$

By substituting $I_U$ in equation (28), we directly derive

$$G = c^2 \frac{R_U}{M_U}.\,^{18} \tag{30}$$

Hence, transactional gravity is able to define $G$ by entropic considerations directly in terms of the total mass in the causal universe. Obviously, equation (30) also offers an understanding of Einstein's formula for the total rest-energy. For a body with rest-mass $m_0$ there holds

$$E_0 = m_0 c^2 = G \frac{m_0 M_U}{R_U}. \tag{31}$$

This way, the total rest energy of a body is equal to the energy stemming from the gravitational potential of the causal universe.

---

[17] Remember that Einstein's equations can be derived as indicated in paragraph 2.4. and shown in [1,2].

[18] If $H_0 = const.$, then (30) turns into $G = \frac{c^3}{H_0 M_U}$.

## 4. Summary and conclusion

Transactional gravity offers deeper reasons for so far merely empirical properties of gravity. The theory explains gravity's universality, the impossibility of screening it off and its propagation by the speed of light. In addition to explaining the existence of a cosmological constant, it confirms the clock hypothesis and explains why the "charge" of gravity is inertial mass. What does it contribute to the reverse question, namely whether inertia is the result of gravity, as Einstein suggested?

First, its relational structure rules out the physical reality of empty space and consequently the existence of absolute space as a reference for acceleration. It further naturally leads to a model for Mach's principle, namely Sciama's origin of inertia. Sciama's theory hinges, next to its Maxwellian structure, on the definition of the potential $\Phi_U$, whose mathematical form can be derived from the findings of transactional gravity. Sciama's model can also be derived from the equations of general relativity [18] which are, in turn, a consequence of transactional gravity. Finally, the theory allows to directly derive the relations (28), (30) from its entropic origin and hence to derive the gravitational constant in terms of the total mass in the causal universe.

By all the above arguments, we realize that, if transactional gravity is a true theory of nature, then general relativity as the field theory of gravity, derived from the entropic consequences of transactions, together with the consequent exclusion of solutions which cannot be realized in nature[19] and the capacity to calculate $G$ in a Machian way, might indeed be the theory which ultimately establishes Mach's view as a true principle of nature.

---

[19] Like vacuum solutions, asymptotically flat solutions or others like in [23].